\documentclass[12pt,a4paper]{article}
\usepackage{amsmath}
\usepackage{latexsym}
\makeatletter
\def\rddots{\mathinner{\mkern1mu\raise\p@%
    \vbox{\kern7\p@\hbox{.}}\mkern2mu%
    \raise4\p@\hbox{.}\mkern2mu\raise7\p@\hbox{.}\mkern1mu}}
\makeatother
\makeatletter
   
   \@addtoreset{equation}{section}
\makeatother
\newcommand{\ket}[1]{{\vert{#1}\rangle}}
\newcommand{\bra}[1]{{\langle{#1}\vert}}

\newcommand{\fukuso}{{\mathbf C}}

\begin{document}

\title{\sl Quantum Damped Harmonic Oscillator}
\author{
  Kazuyuki FUJII
  \thanks{E-mail address : fujii@yokohama-cu.ac.jp }\\
  International College of Arts and Sciences\\
  Yokohama City University\\
  Yokohama, 236--0027\\
  Japan
  }
\date{}
\maketitle

\begin{abstract}
In this chapter we treat the quantum damped harmonic oscillator, and study
mathematical structure of the model, and construct general solution with
any initial condition, and give a quantum counterpart in the case of taking
coherent state as an initial condition.

This is a simple and good model of Quantum Mechanics with dissipation
which is important to understand real world, and readers will get a powerful
weapon for Quantum Physics.
\end{abstract}

\section{Introduction}
In this chapter we introduce a toy model of Quantum Mechanics 
with Dissipation. Quantum Mechanics with Dissipation plays 
a crucial role to understand real world. 
However, it is not easy to master the theory for undergraduates. 
The target of this chapter is eager undergraduates in the world. 
Therefore, a good toy model to understand it deeply is required.

The quantum damped harmonic oscillator is just such a one 
because undergraduates must use (master) many fundamental 
techniques in Quantum Mechanics and Mathematics. That is, 
harmonic oscillator, density operator, Lindblad form, coherent 
state, squeezed state, tensor product, Lie algebra, representation 
theory, Baker--Campbell--Hausdorff formula, etc.

They are ``jewels" in Quantum Mechanics and Mathematics. 
If undergraduates master this model, they will get a powerful 
weapon for Quantum Physics. I expect some of them will 
attack many hard problems of Quantum Mechanics with Dissipation. 

The contents of this chapter are based on our two papers 
\cite{EFS} and \cite{FS}. I will give a clear and fruitful 
explanation to them as much as I can.

\vspace{3mm}
\begin{center}
$\cdots$ {\bf From Yokohama with Love} $\cdots$
\end{center}

\section{Some Preliminaries}
In this section let us make some reviews from Physics and 
Mathematics within our necessity.

\subsection{From Physics}
First we review the solution of classical damped harmonic 
oscillator, which is important to understand the text. 
For this topic see any textbook of Mathematical Physics.

The differential equation is given by
\begin{equation}
\label{eq:classical damped harmonic oscillator}
\ddot{x}+\omega^{2}x=-\gamma \dot{x}
\
\Longleftrightarrow 
\
\ddot{x}+\gamma \dot{x}+\omega^{2}x=0
\quad
(\gamma > 0)
\end{equation}
where $x=x(t),\ \dot{x}=dx/dt$ and the mass $m$ is set to 1 
for simplicity. In the following we treat only the case 
$\omega > \gamma/2$ (the case $\omega=\gamma/2$ may 
be interesting). 

Solutions (with complex form) are well--known to be
\[
x_{\pm}(t)=e^{-\left(\frac{\gamma}{2}\pm i
\sqrt{\omega^{2}-(\frac{\gamma}{2})^{2}}\right)t},
\]
so the general solution is given by
\begin{eqnarray}
\label{eq:general solution}
x(t)=\alpha x_{+}(t)+\bar{\alpha}x_{-}(t)
&=&
\alpha e^{-\left(\frac{\gamma}{2}+i
\sqrt{\omega^{2}-(\frac{\gamma}{2})^{2}}\right)t}
+
\bar{\alpha}e^{-\left(\frac{\gamma}{2}-i
\sqrt{\omega^{2}-(\frac{\gamma}{2})^{2}}\right)t}  \nonumber \\
&=&
\alpha e^{-\left(\frac{\gamma}{2}+i\omega
\sqrt{1-(\frac{\gamma}{2\omega})^{2}}\right)t}
+
\bar{\alpha}e^{-\left(\frac{\gamma}{2}-i\omega
\sqrt{1-(\frac{\gamma}{2\omega})^{2}}\right)t}
\end{eqnarray}
where $\alpha$ is a complex number. 
If $\gamma/2\omega$ is small enough we have 
 an approximate solution
\begin{equation}
\label{eq:approximate solution}
x(t)
\approx 
\alpha e^{-\left(\frac{\gamma}{2}+i\omega\right)t}
+
\bar{\alpha} e^{-\left(\frac{\gamma}{2}-i\omega\right)t}.
\end{equation}

\vspace{5mm}
Next, we consider the quantum harmonic oscillator. This is 
well--known in textbooks of Quantum Mechanics. 
As standard textbooks of Quantum Mechanics see \cite{PD} 
and \cite{HS} (\cite{PD} is particularly interesting).

For the Hamiltonian
\begin{equation}
\label{eq:harmonic oscillator classical}
H=H(q,p)=\frac{1}{2}(p^{2}+\omega^{2}q^{2})
\end{equation}
where $q=q(t),\ p=p(t)$, the canonical equation of motion reads
\[
\dot{q}\equiv\frac{\partial H}{\partial p}=p,\quad
\dot{p}\equiv-\frac{\partial H}{\partial q}=-\omega^{2}q.
\]
From these we recover the equation
\[
\ddot{q}=-\omega^{2}q\ \Longleftrightarrow 
\ddot{q}+\omega^{2}q=0.
\]
See (\ref{eq:classical damped harmonic oscillator}) with $q=x$
and $\lambda=0$.

Next, we introduce the Poisson bracket. For $A=A(q,p),\ B=B(q,p)$ 
it is defined as
\begin{equation}
\label{eq:poisson}
\{A,B\}_{c}\equiv \frac{\partial A}{\partial q}\frac{\partial B}{\partial p}-
\frac{\partial A}{\partial p}\frac{\partial B}{\partial q}
\end{equation}
where $\{,\}_{c}$ means classical. Then it is easy to see
\begin{equation}
\label{eq:classical condition}
\{q,q\}_{c}=0,\quad \{p,p\}_{c}=0,\quad \{q,p\}_{c}=1.
\end{equation}

\vspace{5mm}
Now, we are in a position to give a quantization condition due to 
Dirac. Before that we prepare some notation from algebra. 

Square matrices $A$ and $B$ don't commute in general, so 
we need the commutator
\[
[A,B]=AB-BA.
\]
Then Dirac gives an abstract correspondence 
\quad
$q\longrightarrow \hat{q},\quad p\longrightarrow \hat{p}$ 
\quad which satisfies the condition
\begin{equation}
\label{eq:quantum condition}
[\hat{q},\hat{q}]=0,\quad
[\hat{p},\hat{p}]=0,\quad
[\hat{q},\hat{p}]=i\hbar{\bf 1}
\end{equation}
corresponding to (\ref{eq:classical condition}). 
Here $\hbar$ is the Plank constant, and 
$\hat{q}$ and $\hat{p}$ are both Hermite operators on some 
Fock space (a kind of Hilbert space) given in the latter 
and ${\bf 1}$ is the identity on it. 
Therefore, our quantum Hamiltonian should be
\begin{equation}
\label{eq:harmonic oscillator quantum}
H=H(\hat{q},\hat{p})=
\frac{1}{2}(\hat{p}^{2}+\omega^{2}\hat{q}^{2})
\end{equation}
from (\ref{eq:harmonic oscillator classical}). Note that a notation 
$H$ instead of  $\hat{H}$ is used for simplicity. From now on we 
consider a complex version. From (\ref{eq:harmonic oscillator classical}) 
and (\ref{eq:harmonic oscillator quantum}) we rewrite like
\[
H(q,p)
=\frac{1}{2}(p^{2}+\omega^{2}q^{2})
=\frac{\omega^{2}}{2}(q^{2}+\frac{1}{\omega^{2}}p^{2})
=\frac{\omega^{2}}{2}(q-\frac{i}{\omega}p)(q+\frac{i}{\omega}p)
\]
and 
\begin{eqnarray*}
H(\hat{q},\hat{p})
&=&\frac{\omega^{2}}{2}(\hat{q}^{2}+\frac{1}{\omega^{2}}\hat{p}^{2})
=\frac{\omega^{2}}{2}
\left\{
(\hat{q}-\frac{i}{\omega}\hat{p})(\hat{q}+\frac{i}{\omega}\hat{p})
-\frac{i}{\omega}[\hat{q},\hat{p}]
\right\} \\
&=&\frac{\omega^{2}}{2}
\left\{
(\hat{q}-\frac{i}{\omega}\hat{p})(\hat{q}+\frac{i}{\omega}\hat{p})
+\frac{\hbar }{\omega}
\right\}
=\omega\hbar 
\left\{
\frac{\omega}{2\hbar}(\hat{q}-\frac{i}{\omega}\hat{p})(\hat{q}+\frac{i}{\omega}\hat{p})
+\frac{1}{2}
\right\}
\end{eqnarray*}
by use of (\ref{eq:quantum condition}), and if we set
\begin{equation}
\label{eq:creation-annhilation}
a^{\dagger}=\sqrt{\frac{\omega}{2\hbar}}(\hat{q}-\frac{i}{\omega}\hat{p}),\quad
a=\sqrt{\frac{\omega}{2\hbar}}(\hat{q}+\frac{i}{\omega}\hat{p})
\end{equation}
we have easily
\[
[a,a^{\dagger}]
=\frac{\omega}{2\hbar}
[\hat{q}+\frac{i}{\omega}\hat{p},\hat{q}-\frac{i}{\omega}\hat{p}]
=\frac{\omega}{2\hbar}\left\{-\frac{2i}{\omega}[\hat{q},\hat{p}]\right\}
=\frac{\omega}{2\hbar}\left\{-\frac{2i}{\omega}\times i\hbar \right\}
={\bf 1}
\]
by use of (\ref{eq:quantum condition}). As a result we obtain 
a well--known form
\begin{equation}
\label{eq:quantum hamiltonian}
H=\omega\hbar(a^{\dagger}a+\frac{1}{2}),\quad [a,a^{\dagger}]={\bf 1}.
\end{equation}
Here we used an abbreviation $1/2$ in place of $(1/2){\bf 1}$ for simplicity. 

If we define an operator $N=a^{\dagger}a$ (which is called 
the number operator) then it is easy to see the relations
\begin{equation}
\label{eq:Heisenberg algebra}
[N,a^{\dagger}]=a^{\dagger},\quad [N,a]=-a,\quad [a,a^{\dagger}]={\bf 1}.
\end{equation}
For the proof a well--known formula $[AB,C]=[A,C]B+A[B,C]$ is used. 
Note that $aa^{\dagger}=a^{\dagger}a+[a,a^{\dagger}]=N+1$. 
The set $\{a^{\dagger},a,N\}$ is just a generator of Heisenberg algebra 
and we can construct a Fock space based on this.  Let us note that 
$a$, $a^{\dagger}$ and $N$ are called the annihilation operator, 
creation one and number one respectively.

\vspace{5mm}
First of all let us define a vacuum $|0\rangle$. This is defined by 
the equation $a|0\rangle=0$. Based on this vacuum we construct 
the $n$ state $|n\rangle$ like
\[
|n\rangle=\frac{(a^{\dagger})^{n}}{\sqrt{n!}}|0\rangle
\quad (0\leq n).
\]
Then we can easily prove
\begin{equation}
\label{eq:fundamental equations}
a^{\dagger}|n\rangle=\sqrt{n+1}|n+1\rangle, \quad
a|n\rangle=\sqrt{n}|n-1\rangle, \quad
N|n\rangle=n|n\rangle
\end{equation}
and moreover can prove both the orthogonality condition and 
the resolution of unity 
\begin{equation}
\label{eq:resolution of unity}
\langle m|n\rangle=\delta_{mn},\quad
\sum_{n=0}^{\infty}|n\rangle \langle n|={\bf 1}.
\end{equation}
For the proof one can use for example
\[
a^{2}(a^{\dagger})^{2}=a(aa^{\dagger})a^{\dagger}=a(N+1)a^{\dagger}=
(N+2)aa^{\dagger}=(N+2)(N+1)
\]
by (\ref{eq:Heisenberg algebra}), therefore we have
\[
\langle 0|a^{2}(a^{\dagger})^{2}|0 \rangle=
\langle 0|(N+2)(N+1)|0 \rangle = 2!
\Longrightarrow \langle 2|2\rangle=1.
\]
The proof of the resolution of unity may be not easy for readers 
(we omit it here). 

As a result we can define a Fock space generated by the generator 
$\{a^{\dagger},a,N\}$
\begin{equation}
\label{eq:Fock space}
{\cal F}=\left\{\sum_{n=0}^{\infty}c_{n}|n\rangle\ \in\ {\bf C}^{\infty}
\ |\ 
\sum_{n=0}^{\infty}|c_{n}|^{2}<\infty \right\}.
\end{equation}
This is just a kind of Hilbert space. On this space the operators (=
infinite dimensional matrices) $a^{\dagger}$, $a$ and $N$ are 
represented as
\begin{eqnarray}
\label{eq:creation-annihilation}
&&a=
\left(
\begin{array}{ccccc}
0 & 1 &          &          &        \\
  & 0 & \sqrt{2} &          &        \\
  &   & 0        & \sqrt{3} &        \\
  &   &          & 0        & \ddots \\
  &   &          &          & \ddots
\end{array}
\right),\quad
a^{\dagger}=
\left(
\begin{array}{ccccc}
0 &          &          &        &        \\
1 & 0        &          &        &        \\
  & \sqrt{2} & 0        &        &        \\
  &          & \sqrt{3} & 0      &        \\
  &          &          & \ddots & \ddots
\end{array}
\right),
\nonumber \\
&&N=a^{\dagger}a=
\left(
\begin{array}{ccccc}
0 &   &   &   &        \\
  & 1 &   &   &        \\
  &   & 2 &   &        \\
  &   &   & 3 &        \\
  &   &   &   & \ddots
\end{array}
\right)
\end{eqnarray}
by use of (\ref{eq:fundamental equations}).

\vspace{3mm}\noindent
{\bf Note}\ \ We can add a phase to $\{a,a^{\dagger}\}$ like
\[
b=e^{i\theta}a,\quad  b^{\dagger}=e^{-i\theta}a^{\dagger},\quad 
N=b^{\dagger}b=a^{\dagger}a
\]
where $\theta$ is constant. Then we have another Heisenberg algebra
\[
[N,b^{\dagger}]=b^{\dagger},\quad [N,b]=-b,\quad [b,b^{\dagger}]={\bf 1}.
\]

\vspace{3mm}
Next, we introduce a coherent state which plays a central role 
in Quantum Optics or Quantum Computation. For $z\in {\bf C}$ 
the coherent state $|z\rangle\in {\cal F}$ is defined by  
the equation
\[
a|z\rangle=z|z\rangle\quad\mbox{and}\quad \langle{z}|{z}\rangle=1.
\]
The annihilation operator $a$ is not hermitian, so this equation 
is never trivial. For this state the following three equations 
are equivalent :
\begin{equation}
\label{eq:three equations}
\left\{
\begin{array}{lll}
(1)\quad a|z\rangle=z|z\rangle\ \mbox{and}\ \langle{z}|{z}\rangle=1, \\
(2)\quad |z\rangle=e^{za^{\dagger}-\bar{z}a}|0\rangle, \\
(3)\quad |z\rangle=e^{-\frac{|z|^{2}}{2}}\sum_{n=0}^{\infty}\frac{z^{n}}{\sqrt{n!}}|n\rangle.
\end{array}
\right.
\end{equation}
The proof is as follows. From (1) to (2) we use a popular formula 
\[
e^{A}Be^{-A}=B+[A,B]+\frac{1}{2!}[A,[A,B]]+\cdots
\]
($A$, $B$ : operators) to prove
\[
e^{-(za^{\dagger}-\bar{z}a)}ae^{za^{\dagger}-\bar{z}a}=a+z.
\]
From (2) to (3) we use the Baker-Campbell-Hausdorff formula (see 
for example \cite{CZ})
\[
e^{A}e^{B}=e^{A+B+\frac{1}{2}[A,B]+\frac{1}{6}[A,[A,B]]+\frac{1}{6}[B,[A,B]]+\cdots}.
\]
If $[A,[A,B]]=0=[B,[A,B]]$ (namely, $[A,B]$ commutes with both $A$ and $B$) 
then we have
\begin{equation}
\label{eq:special B-C-H}
e^{A}e^{B}=e^{A+B+\frac{1}{2}[A,B]}=e^{\frac{1}{2}[A,B]}e^{A+B}
\Longrightarrow e^{A+B}=e^{-\frac{1}{2}[A,B]}e^{A}e^{B}.
\end{equation}
In our case the condition is satisfied because of $[a,a^{\dagger}]=1$. 
Therefore we obtain a (famous) decomposition
\begin{equation}
\label{eq:famous decomposition}
e^{za^{\dagger}-\bar{z}a}=e^{-\frac{|z|^{2}}{2}}e^{za^{\dagger}}e^{-\bar{z}a}.
\end{equation}
The remaining part of the proof is left to readers.

From the equation (3) in (\ref{eq:three equations}) 
we obtain the resolution of unity for coherent states
\begin{equation}
\int\int\frac{dxdy}{\pi}|{z}\rangle\langle{z}|
=\sum_{n=0}^{\infty}|n\rangle \langle n|={\bf 1}\quad (z=x+iy).
\end{equation}
The proof is reduced to the following formula
\[
\int\int\frac{dxdy}{\pi}e^{-|z|^{2}}{\bar{z}}^{m}z^{n}=
n!\ \delta_{mn}\quad (z=x+iy).
\]
The proof is left to readers.  See \cite{KS} for more 
general knowledge of coherent states.

\subsection{From Mathematics}
We consider a simple matrix equation
\begin{equation}
\label{eq:matrix equation}
\frac{d}{dt}X=AXB
\end{equation}
where
\[
X=X(t)=
\left(
\begin{array}{cc}
x_{11}(t) & x_{12}(t) \\
x_{21}(t) & x_{22}(t)
\end{array}
\right),\quad
A=
\left(
\begin{array}{cc}
a_{11} & a_{12} \\
a_{21} & a_{22}
\end{array}
\right),\quad
B=
\left(
\begin{array}{cc}
b_{11} & b_{12} \\
b_{21} & b_{22}
\end{array}
\right).
\]
A standard form of linear differential equation which we usually 
treat is
\[
\frac{d}{dt}{\bf x}=C{\bf x}
\]
where ${\bf x}={\bf x}(t)$ is a vector and $C$ is a matrix associated 
to the vector. 
Therefore, we want to rewrite (\ref{eq:matrix equation}) into 
a standard form.

For the purpose we introduce the Kronecker product of matrices. 
For example, it is defined as
\begin{eqnarray}
\label{eq:Kronecker product}
A\otimes B
&=&
\left(
\begin{array}{cc}
a_{11} & a_{12} \\
a_{21} & a_{22}
\end{array}
\right)\otimes B
\equiv 
\left(
\begin{array}{cc}
a_{11}B & a_{12}B \\
a_{21}B & a_{22}B
\end{array}
\right)  \nonumber \\
&=&
\left(
  \begin{array}{cccc}
    a_{11}b_{11} & a_{11}b_{12} & a_{12}b_{11} & a_{12}b_{12}  \\
    a_{11}b_{21} & a_{11}b_{22} & a_{12}b_{21} & a_{12}b_{22}  \\
    a_{21}b_{11} & a_{21}b_{12} & a_{22}b_{11} & a_{22}b_{12}  \\
    a_{21}b_{21} & a_{21}b_{22} & a_{22}b_{21} & a_{22}b_{22}    
  \end{array}
\right)
\end{eqnarray}
for $A$ and $B$ above. Note that recently we use the tensor 
product instead of the Kronecker product, so we use it 
in the following. 
Here, let us list some useful properties of the tensor product
\begin{eqnarray}
\label{eq:useful properties}
&&(1)\quad (A_{1}\otimes B_{1})(A_{2}\otimes B_{2})=A_{1}A_{2}\otimes B_{1}B_{2}, 
\nonumber \\
&&(2)\quad (A\otimes E)(E\otimes B)=A\otimes B=(E\otimes B)(A\otimes E), 
\nonumber \\
&&(3)\quad e^{A\otimes E+E\otimes B}=e^{A\otimes E}e^{E\otimes B}=
        (e^{A}\otimes E)(E\otimes e^{B})=e^{A}\otimes e^{B},  \\
&&(4)\quad (A\otimes B)^{\dagger}=A^{\dagger}\otimes B^{\dagger} \nonumber
\end{eqnarray}
where $E$ is the unit matrix. The proof is left to readers. \cite{Five} is 
recommended as a general introduction.

Then the equation (\ref{eq:matrix equation}) can be written in terms of 
components as
\[
\left\{
\begin{array}{llll}
\frac{dx_{11}}{dt}&=&a_{11}b_{11}x_{11}+a_{11}b_{21}x_{12}+a_{12}b_{11}x_{21}+a_{12}b_{21}x_{22}, \\
\frac{dx_{12}}{dt}&=&a_{11}b_{12}x_{11}+a_{11}b_{22}x_{12}+a_{12}b_{12}x_{21}+a_{12}b_{22}x_{22}, \\
\frac{dx_{21}}{dt}&=&a_{21}b_{11}x_{11}+a_{21}b_{21}x_{12}+a_{22}b_{11}x_{21}+a_{22}b_{21}x_{22}, \\
\frac{dx_{22}}{dt}&=&a_{21}b_{12}x_{11}+a_{21}b_{22}x_{12}+a_{22}b_{12}x_{21}+a_{22}b_{22}x_{22}
\end{array}
\right.
\]
or in a matrix form
\[
\frac{d}{dt}
\left(
\begin{array}{l}
x_{11} \\
x_{12} \\
x_{21} \\
x_{22}
\end{array}
\right)
=
\left(
\begin{array}{cccc}
a_{11}b_{11} & a_{11}b_{21} & a_{12}b_{11} & a_{12}b_{21} \\
a_{11}b_{12} & a_{11}b_{22} & a_{12}b_{12} & a_{12}b_{22} \\
a_{21}b_{11} & a_{21}b_{21} & a_{22}b_{11} & a_{22}b_{21} \\
a_{21}b_{12} & a_{21}b_{22} & a_{22}b_{12} & a_{22}b_{22}
\end{array}
\right)
\left(
\begin{array}{l}
x_{11} \\
x_{12} \\
x_{21} \\
x_{22}
\end{array}
\right).
\]
If we set 
\[
X=
\left(
\begin{array}{cc}
x_{11} & x_{12} \\
x_{21} & x_{22}
\end{array}
\right)
\ \Longrightarrow\ \widehat{X}=(x_{11},x_{12},x_{21},x_{22})^{T}
\]
where $T$ is the transpose, then we obtain a standard form
\begin{equation}
\label{eq:axb}
\frac{d}{dt}X=AXB\ \Longrightarrow\  
\frac{d}{dt}\widehat{X}=(A\otimes B^{T})\widehat{X}
\end{equation}
from (\ref{eq:Kronecker product}).

Similarly we have a standard form
\begin{equation}
\label{eq:ax+xb}
\frac{d}{dt}X=AX+XB\ \Longrightarrow\  
\frac{d}{dt}\widehat{X}=(A\otimes E+E\otimes B^{T})\widehat{X}
\end{equation}
where $E^{T}=E$ for the unit matrix $E$.

From these lessons there is no problem to generalize 
(\ref{eq:axb}) and (\ref{eq:ax+xb}) based on $2\times 2$ 
matrices to ones based on any (square) matrices or operators 
on ${\cal F}$. Namely, we have
\begin{equation}
\label{eq:standard form}
\left\{
\begin{array}{ll}
\frac{d}{dt}X=AXB\ \Longrightarrow\  
\frac{d}{dt}\widehat{X}=(A\otimes B^{T})\widehat{X}, \\
\frac{d}{dt}X=AX+XB\ \Longrightarrow\  
\frac{d}{dt}\widehat{X}=(A\otimes I+I\otimes B^{T})\widehat{X}.
\end{array}
\right.
\end{equation}
where $I$ is the identity $E$ (matrices) or ${\bf 1}$ (operators).

\section{Quantum Damped Harmonic Oscillator}
In this section we treat the quantum damped harmonic oscillator. 
As a general introduction to this topic see \cite{BP} or \cite{WS}.

\subsection{Model}
Before that we introduce the quantum harmonic oscillator. 
The Schr\"{o}dinger equation is given by
\[
i\hbar\frac{\partial}{\partial t}|\Psi(t)\rangle
=H|\Psi(t)\rangle
=\left(\omega\hbar(N+\frac{1}{2})\right)|\Psi(t)\rangle
\]
by (\ref{eq:quantum hamiltonian}) (note $N=a^{\dagger}a$). 
In the following we use $\frac{\partial}{\partial t}$ instead of 
$\frac{d}{dt}$.

Now we change from a wave--function to a density operator 
because we want to treat a mixed state, which is a well--known 
technique in Quantum Mechanics or Quantum Optics.

If we set $\rho(t)=|\Psi(t)\rangle\langle\Psi(t)|$, then a little algebra 
gives
\begin{equation}
\label{eq:Liouville equation}
i\hbar\frac{\partial}{\partial t}\rho=[H,\rho]=[\omega\hbar N,\rho]
\ \Longrightarrow\ 
\frac{\partial}{\partial t}\rho=-i[\omega N,\rho].
\end{equation}
This is called the quantum Liouville equation. 
With this form we can treat a mixed state like
\[
\rho=\rho(t)=\sum_{j=1}^{N}u_{j}|\Psi_{j}(t)\rangle\langle\Psi_{j}(t)|
\]
where $u_{j}\geq 0$ and $\sum_{j=1}^{N}u_{j}=1$. 
Note that the general solution of (\ref{eq:Liouville equation}) 
is given by
\[
\rho(t)=e^{-i\omega{N}t}\rho(0)e^{i\omega{N}t}.
\]

We are in a position to state the equation of 
quantum damped harmonic oscillator by use of  
(\ref{eq:Liouville equation}).

\vspace{3mm}\noindent
{\bf Definition} The equation is given by
\begin{equation}
\label{eq:quantum damped harmonic oscillator}
\frac{\partial}{\partial t}\rho=-i[\omega a^{\dagger}a,\rho]
-\frac{\mu}{2}
\left(a^{\dagger}a\rho+\rho a^{\dagger}a-2a\rho a^{\dagger}\right)
-\frac{\nu}{2}
\left(aa^{\dagger}\rho+\rho aa^{\dagger}-2a^{\dagger}\rho{a}\right)
\end{equation}
where $\mu,\ \nu$ ($\mu > \nu \geq 0$) are some real constants 
depending on the system (for example, a damping rate of the cavity 
mode)\footnote{The aim of this chapter is not to drive this 
equation. In fact, its derivation is not easy for non--experts, so 
see for example the original papers \cite{Lind} and \cite{GKS}, or 
\cite{KH} as a short review paper}.

\vspace{3mm}
Note that the extra term
\[
-\frac{\mu}{2}
\left(a^{\dagger}a\rho+\rho a^{\dagger}a-2a\rho a^{\dagger}\right)
-\frac{\nu}{2}
\left(aa^{\dagger}\rho+\rho aa^{\dagger}-2a^{\dagger}\rho{a}\right)
\]
is called the Lindblad form (term). Such a term plays an essential 
role in {\bf Decoherence}.

\subsection{Method of Solution}
First we solve the Lindblad equation :
\begin{equation}
\label{eq:Lindblad form}
\frac{\partial}{\partial t}\rho=
-\frac{\mu}{2}
\left(a^{\dagger}a\rho+\rho a^{\dagger}a-2a\rho a^{\dagger}\right)
-\frac{\nu}{2}
\left(aa^{\dagger}\rho+\rho aa^{\dagger}-2a^{\dagger}\rho{a}\right).
\end{equation}
Interesting enough, we can solve this equation completely.

Let us rewrite (\ref{eq:Lindblad form}) more conveniently using 
the number operator $N\equiv a^{\dagger}a$
\begin{equation}
\label{eq:Lindblad form-2}
\frac{\partial}{\partial t}\rho
=
\mu a\rho a^{\dagger}+\nu a^{\dagger}\rho{a}
-\frac{\mu+\nu}{2}(N\rho+\rho N +\rho)+\frac{\mu-\nu}{2}\rho
\end{equation}
where we have used $aa^{\dagger}=N+{\bf 1}$.

From here we use the method developed in Section 2.2. 
For a matrix $X=(x_{ij})\in M({\cal F})$ over ${\cal F}$
\[
X=
\left(
\begin{array}{cccc}
x_{00} & x_{01} & x_{02} & \cdots  \\
x_{10} & x_{11} & x_{12} & \cdots  \\
x_{20} & x_{21} & x_{22} & \cdots  \\
\vdots & \vdots & \vdots & \ddots
\end{array}
\right)
\]
we correspond to the vector $\widehat{X}\in 
{{\cal F}}^{\mbox{dim}_{\fukuso}{\cal F}}$ as
\begin{equation}
\label{eq:correspondence}
X=(x_{ij})\ \longrightarrow\ 
\widehat{X}=(x_{00},x_{01},x_{02},\cdots;x_{10},x_{11},x_{12},\cdots;
x_{20},x_{21},x_{22},\cdots;\cdots \cdots)^{T}
\end{equation}
where $T$ means the transpose. The following formulas
\begin{equation}
\label{eq:well-known formulas}
\widehat{AXB}=(A\otimes B^{T})\widehat{X},\quad
\widehat{(AX+XB)}=(A\otimes {\bf 1}+{\bf 1}\otimes B^{T})\widehat{X}
\end{equation}
hold for $A,B,X\in M({\cal F})$, see (\ref{eq:standard form}).

Then (\ref{eq:Lindblad form-2}) becomes
\begin{eqnarray}
\label{eq:rho-equation}
\frac{\partial}{\partial t}\widehat{\rho}
&=&
\left\{
\mu a\otimes (a^{\dagger})^{T}+\nu a^{\dagger}\otimes a^{T}
-\frac{\mu+\nu}{2}(N\otimes {\bf 1}+{\bf 1}\otimes N+{\bf 1}\otimes {\bf 1})
+\frac{\mu-\nu}{2}{\bf 1}\otimes {\bf 1}
\right\}
\widehat{\rho} \nonumber \\
&=&
\left\{
\frac{\mu-\nu}{2}{\bf 1}\otimes {\bf 1}+
\nu a^{\dagger}\otimes a^{\dagger}+\mu a\otimes a
-\frac{\mu+\nu}{2}(N\otimes {\bf 1}+{\bf 1}\otimes N+{\bf 1}\otimes {\bf 1})
\right\}
\widehat{\rho}
\end{eqnarray}
where we have used $a^{T}=a^{\dagger}$ from the form 
(\ref{eq:creation-annihilation}), so that the solution is formally given by
\begin{equation}
\label{eq:formal-solution}
\widehat{\rho}(t)=
e^{\frac{\mu-\nu}{2}t}
e^
{t
\left\{
\nu a^{\dagger}\otimes a^{\dagger}+\mu a\otimes a
-\frac{\mu+\nu}{2}(N\otimes {\bf 1}+{\bf 1}\otimes N+{\bf 1}\otimes {\bf 1})
\right\}
}
\widehat{\rho}(0).
\end{equation}

In order to use some techniques from Lie algebra we set
\begin{equation}
\label{eq:K-generators}
K_{3}=\frac{1}{2}(N\otimes {\bf 1}+{\bf 1}\otimes N+{\bf 1}\otimes {\bf 1}),
\quad
K_{+}=a^{\dagger}\otimes a^{\dagger},\quad
K_{-}=a\otimes a \quad \left(K_{-}=K_{+}^{\dagger}\right)
\end{equation}
then we can show the relations
\[
[K_{3},K_{+}]=K_{+},\quad [K_{3},K_{-}]=-K_{-},\quad [K_{+},K_{-}]=-2K_{3}.
\]
This is just the $su(1,1)$ algebra. The proof is very easy and is left 
to readers. 

The equation (\ref{eq:formal-solution}) can be written simply as
\begin{equation}
\label{eq:formal-solution-2}
\widehat{\rho}(t)=e^{\frac{\mu-\nu}{2}t}
e^{t\{\nu K_{+}+\mu K_{-}-(\mu+\nu)K_{3}\}}\widehat{\rho}(0),
\end{equation}
so we have only to calculate the term
\begin{equation}
\label{eq:exponential}
e^{t\{\nu K_{+}+\mu K_{-}-(\mu+\nu)K_{3}\}},
\end{equation}
which is of course not simple. Now the disentangling formula in \cite{KF-1} is 
helpful in calculating (\ref{eq:exponential}). 

If we set $\{k_{+},k_{-},k_{3}\}$ as
\begin{equation}
\label{eq:su(1,1) generators}
 k_{+} = \left(
        \begin{array}{cc}
               0 & 1 \\
               0 & 0 
        \end{array}
       \right),
 \quad 
 k_{-} = \left(
        \begin{array}{cc}
               0 & 0 \\
              -1 & 0 
        \end{array}
       \right),
 \quad 
 k_{3} = \frac12 
       \left(
        \begin{array}{cc}
               1 &  0 \\
               0 & -1 
        \end{array}
       \right)
 \quad
 \left(k_{-}\ne k_{+}^{\dagger}\right)
\end{equation}
then it is very easy to check the relations
\[
[k_{3},k_{+}]=k_{+},\quad [k_{3},k_{-}]=-k_{-},\quad 
[k_{+},k_{-}]=-2k_{3}.
\]
That is, $\{k_{+},k_{-},k_{3}\}$ are generators of the Lie algebra 
$su(1,1)$. Let us show by $SU(1,1)$ the corresponding Lie 
group, which is a typical noncompact group.

Since $SU(1,1)$ is contained in the special linear group $SL(2;\fukuso)$, 
we {\bf assume} that there exists an infinite dimensional unitary 
representation 
$\rho : SL(2;\fukuso)\ \longrightarrow\ U({\cal F}\otimes {\cal F})$ 
(group homomorphism) 
satisfying
\[
d\rho(k_{+})=K_{+},\quad d\rho(k_{-})=K_{-},\quad d\rho(k_{3})=K_{3}.
\]

From (\ref{eq:exponential}) some algebra gives
\begin{eqnarray}
\label{eq:assumption}
e^{t\{\nu K_{+}+\mu K_{-}-(\mu+\nu)K_{3}\}}
&=&
e^{t\{\nu d\rho(k_{+})+\mu d\rho(k_{-})-(\mu+\nu)d\rho(k_{3})\}}
\nonumber \\
&=&
e^{d\rho(t(\nu k_{+}+\mu k_{-}-(\mu+\nu) k_{3}))}  \nonumber \\
&=&
\rho\left(e^{t(\nu k_{+}+\mu k_{-}-(\mu+\nu) k_{3})}\right) 
\quad (\Downarrow\ \mbox{by definition}) \nonumber \\
&\equiv& 
\rho\left(e^{tA}\right)
\end{eqnarray}
and we have
\begin{eqnarray*}
e^{tA}
&=&
e^{t\left\{\nu k_{+}+\mu k_{-}-(\mu+\nu)k_{3}\right\}} \\
&=&
\exp{
\left\{t
  \left(
   \begin{array}{cc}
    -\frac{\mu+\nu}{2} & \nu \\
    -\mu & \frac{\mu+\nu}{2}
   \end{array}
  \right)
\right\}
} \\
&=&
\left(
 \begin{array}{cc}
  \cosh\left(\frac{\mu-\nu}{2}t\right)-\frac{\mu+\nu}{\mu-\nu}
  \sinh\left(\frac{\mu-\nu}{2}t\right) & 
    \frac{2\nu}{\mu-\nu}\sinh\left(\frac{\mu-\nu}{2}t\right)  \\
  -\frac{2\mu}{\mu-\nu}\sinh\left(\frac{\mu-\nu}{2}t\right) &
  \cosh\left(\frac{\mu-\nu}{2}t\right)+\frac{\mu+\nu}{\mu-\nu}
  \sinh\left(\frac{\mu-\nu}{2}t\right)
 \end{array}
\right).
\end{eqnarray*}
The proof is based on the following two facts.
\[
(tA)^{2}=t^{2}
  \left(
   \begin{array}{cc}
    -\frac{\mu+\nu}{2} & \nu \\
    -\mu & \frac{\mu+\nu}{2}
   \end{array}
  \right)^{2}
=t^{2}
  \left(
   \begin{array}{cc}
    \left(\frac{\mu+\nu}{2}\right)^{2}-\mu\nu & 0 \\
    0 & \left(\frac{\mu+\nu}{2}\right)^{2}-\mu\nu
   \end{array}
  \right)
=\left(\frac{\mu-\nu}{2}t\right)^{2}
  \left(
  \begin{array}{cc}
  1 & 0 \\
  0 & 1
  \end{array}
  \right)
\]
and
\[
e^{X}
=\sum_{n=0}^{\infty}\frac{1}{n!}X^{n}
=\sum_{n=0}^{\infty}\frac{1}{(2n)!}X^{2n}+\sum_{n=0}^{\infty}\frac{1}{(2n+1)!}X^{2n+1}
\quad (X=tA).
\]
Note that
\[
\cosh(x)=\frac{e^{x}+e^{-x}}{2}=\sum_{n=0}^{\infty}\frac{x^{2n}}{(2n)!}
\quad \mbox{and}\quad
\sinh(x)=\frac{e^{x}-e^{-x}}{2}=\sum_{n=0}^{\infty}\frac{x^{2n+1}}{(2n+1)!}.
\]
The remainder is left to readers.

The Gauss decomposition formula (in $SL(2;{\bf C})$)
\[
  \left(
   \begin{array}{cc}
     a & b \\
     c & d \\
   \end{array}
  \right)
=
  \left(
   \begin{array}{cc}
     1 & \frac{b}{d} \\
     0 & 1
   \end{array}
  \right)
  \left(
   \begin{array}{cc}
     \frac{1}{d} & 0 \\
     0 & d
   \end{array}
  \right)
  \left(
   \begin{array}{cc}
     1 & 0            \\
     \frac{c}{d} & 1
   \end{array}
  \right)\quad (ad-bc=1)
\]
gives the decomposition

\begin{eqnarray*}
e^{tA}
=\hspace{-5mm}
  &&\left(
   \begin{array}{cc}
     1 & \frac{\frac{2\nu}{\mu-\nu}\sinh\left(\frac{\mu-\nu}{2}t\right)}
     {\cosh\left(\frac{\mu-\nu}{2}t\right)+\frac{\mu+\nu}{\mu-\nu}
      \sinh\left(\frac{\mu-\nu}{2}t\right)} \\
     0 & 1
   \end{array}
  \right)\times     \\
  &&\left(
   \begin{array}{cc}
     \frac{1}{\cosh\left(\frac{\mu-\nu}{2}t\right)+\frac{\mu+\nu}{\mu-\nu}
              \sinh\left(\frac{\mu-\nu}{2}t\right)} & 0 \\
     0 & \cosh\left(\frac{\mu-\nu}{2}t\right)+\frac{\mu+\nu}{\mu-\nu}
         \sinh\left(\frac{\mu-\nu}{2}t\right)
   \end{array}
  \right)\times     \\
  &&\left(
   \begin{array}{cc}
     1 & 0            \\
     -\frac{\frac{2\mu}{\mu-\nu}\sinh\left(\frac{\mu-\nu}{2}t\right)}
     {\cosh\left(\frac{\mu-\nu}{2}t\right)+\frac{\mu+\nu}{\mu-\nu}
      \sinh\left(\frac{\mu-\nu}{2}t\right)} & 1
   \end{array}
  \right)
\end{eqnarray*}
and moreover we have
\begin{eqnarray*}
e^{tA}
=
&\exp&\hspace{-3mm}
  \left(
   \begin{array}{cc}
     0 & \frac{\frac{2\nu}{\mu-\nu}\sinh\left(\frac{\mu-\nu}{2}t\right)}
     {\cosh\left(\frac{\mu-\nu}{2}t\right)+\frac{\mu+\nu}{\mu-\nu}
      \sinh\left(\frac{\mu-\nu}{2}t\right)}  \\
     0 & 0
   \end{array}
  \right)\times   \\
&\exp&\hspace{-3mm}
  \left(
   \begin{array}{cc}
     -\log\left(\cosh\left(\frac{\mu-\nu}{2}t\right)+\frac{\mu+\nu}{\mu-\nu}
                \sinh\left(\frac{\mu-\nu}{2}t\right)\right) & 0 \\
     0 & \log\left(\cosh\left(\frac{\mu-\nu}{2}t\right)+\frac{\mu+\nu}{\mu-\nu}
         \sinh\left(\frac{\mu-\nu}{2}t\right)\right)
   \end{array}
  \right)\times   \\
&\exp&\hspace{-3mm}
  \left(
   \begin{array}{cc}
     0 & 0            \\
     -\frac{\frac{2\mu}{\mu-\nu}\sinh\left(\frac{\mu-\nu}{2}t\right)}
     {\cosh\left(\frac{\mu-\nu}{2}t\right)+\frac{\mu+\nu}{\mu-\nu}
      \sinh\left(\frac{\mu-\nu}{2}t\right)} & 0
   \end{array}
  \right)     \\
=
&\exp&\hspace{-3mm}
   \left(\frac{\frac{2\nu}{\mu-\nu}\sinh\left(\frac{\mu-\nu}{2}t\right)}
     {\cosh\left(\frac{\mu-\nu}{2}t\right)+\frac{\mu+\nu}{\mu-\nu}
      \sinh\left(\frac{\mu-\nu}{2}t\right)}k_{+}\right)\times  \\
&\exp&\hspace{-3mm}
    \left(-2\log\left(\cosh\left(\frac{\mu-\nu}{2}t\right)+
           \frac{\mu+\nu}{\mu-\nu}\sinh\left(\frac{\mu-\nu}{2}t\right)\right)
    k_{3}\right)\times  \\
&\exp&\hspace{-3mm}
     \left(\frac{\frac{2\mu}{\mu-\nu}\sinh\left(\frac{\mu-\nu}{2}t\right)}
     {\cosh\left(\frac{\mu-\nu}{2}t\right)+\frac{\mu+\nu}{\mu-\nu}
      \sinh\left(\frac{\mu-\nu}{2}t\right)}k_{-}\right).
\end{eqnarray*}

Since $\rho$ is a group homomorphism ($\rho(XYZ)=\rho(X)\rho(Y)\rho(Z)$) 
and the formula $\rho\left(e^{Lk}\right)=e^{Ld\rho(k)}$ 
($k=k_{+},k_{3},k_{-}$) holds we obtain
\begin{eqnarray*}
\rho\left(e^{tA}\right)
=\hspace{-5mm}
&&\exp
   \left(\frac{\frac{2\nu}{\mu-\nu}\sinh\left(\frac{\mu-\nu}{2}t\right)}
     {\cosh\left(\frac{\mu-\nu}{2}t\right)+\frac{\mu+\nu}{\mu-\nu}
      \sinh\left(\frac{\mu-\nu}{2}t\right)}d\rho(k_{+})\right)\times  \\
&&\exp
    \left(-2\log\left(\cosh\left(\frac{\mu-\nu}{2}t\right)+
           \frac{\mu+\nu}{\mu-\nu}\sinh\left(\frac{\mu-\nu}{2}t\right)\right)
    d\rho(k_{3})\right)\times  \\
&&\exp
     \left(\frac{\frac{2\mu}{\mu-\nu}\sinh\left(\frac{\mu-\nu}{2}t\right)}
     {\cosh\left(\frac{\mu-\nu}{2}t\right)+\frac{\mu+\nu}{\mu-\nu}
      \sinh\left(\frac{\mu-\nu}{2}t\right)}d\rho(k_{-})\right).
\end{eqnarray*}

As a result we have the disentangling formula
\begin{eqnarray}
\label{eq:our disentangling formula}
e^{t\{\nu K_{+}+\mu K_{-}-(\mu+\nu)K_{3}\}}
=
&\exp&\hspace{-3mm}
   \left(\frac{\frac{2\nu}{\mu-\nu}\sinh\left(\frac{\mu-\nu}{2}t\right)}
     {\cosh\left(\frac{\mu-\nu}{2}t\right)+\frac{\mu+\nu}{\mu-\nu}
      \sinh\left(\frac{\mu-\nu}{2}t\right)}K_{+}\right)\times  \nonumber \\
&\exp&\hspace{-3mm}
    \left(-2\log\left(\cosh\left(\frac{\mu-\nu}{2}t\right)+
           \frac{\mu+\nu}{\mu-\nu}\sinh\left(\frac{\mu-\nu}{2}t\right)\right)
    K_{3}\right)\times  \nonumber \\
&\exp&\hspace{-3mm}
     \left(\frac{\frac{2\mu}{\mu-\nu}\sinh\left(\frac{\mu-\nu}{2}t\right)}
     {\cosh\left(\frac{\mu-\nu}{2}t\right)+\frac{\mu+\nu}{\mu-\nu}
      \sinh\left(\frac{\mu-\nu}{2}t\right)}K_{-}\right)
\end{eqnarray}
by (\ref{eq:assumption}). 

In the following we set for simplicity
\begin{eqnarray}
\label{eq:a set of solutions}
E(t)&=&\frac{\frac{2\mu}{\mu-\nu}\sinh\left(\frac{\mu-\nu}{2}t\right)}
     {\cosh\left(\frac{\mu-\nu}{2}t\right)+\frac{\mu+\nu}{\mu-\nu}
      \sinh\left(\frac{\mu-\nu}{2}t\right)},  \nonumber \\
F(t)&=&\cosh\left(\frac{\mu-\nu}{2}t\right)+
     \frac{\mu+\nu}{\mu-\nu}\sinh\left(\frac{\mu-\nu}{2}t\right),  \\
G(t)&=&\frac{\frac{2\nu}{\mu-\nu}\sinh\left(\frac{\mu-\nu}{2}t\right)}
     {\cosh\left(\frac{\mu-\nu}{2}t\right)+\frac{\mu+\nu}{\mu-\nu}
      \sinh\left(\frac{\mu-\nu}{2}t\right)}.  \nonumber
\end{eqnarray}

\vspace{5mm}
Readers should be careful of this ``proof", which is a heuristic method. 
In fact, it is incomplete because we have assumed a group homomorphism.  
In order to complete it we want to show a disentangling formula like
\[
e^{t\{\nu K_{+}+\mu K_{-}-(\mu+\nu)K_{3}\}}=
e^{f(t)K_{+}}e^{g(t)K_{3}}e^{h(t)K_{-}}
\]
with unknowns $f(t),\ g(t),\ h(t)$ satisfying $f(0)=g(0)=h(0)=0$. 
For the purpose we set
\[
A(t)=e^{t\{\nu K_{+}+\mu K_{-}-(\mu+\nu)K_{3}\}}, \quad
B(t)=e^{f(t)K_{+}}e^{g(t)K_{3}}e^{h(t)K_{-}}.
\]
For $t=0$ we have $A(0)=B(0)=\mbox{identity}$ and 
\[
\dot{A}(t)=\{\nu K_{+}+\mu K_{-}-(\mu+\nu)K_{3}\}A(t).
\]
Next, let us calculate $\dot{B}(t)$. By use of the Leibniz rule
\begin{eqnarray*}
\dot{B}(t)
&=&(\dot{f}K_{+})e^{f(t)K_{+}}e^{g(t)K_{3}}e^{h(t)K_{-}}+
e^{f(t)K_{+}}(\dot{g}K_{3})e^{g(t)K_{3}}e^{h(t)K_{-}}+
e^{f(t)K_{+}}e^{g(t)K_{3}}(\dot{h}K_{-})e^{h(t)K_{-}} \\
&=&\left\{\dot{f}K_{+}+\dot{g}e^{fK_{+}}K_{3}e^{-fK_{+}}+
\dot{h}e^{fK_{+}}e^{gK_{3}}K_{-}e^{-gK_{3}}e^{-fK_{+}}\right\}
e^{f(t)K_{+}}e^{g(t)K_{3}}e^{h(t)K_{-}} \\
&=&\left\{\dot{f}K_{+}+\dot{g}(K_{3}-fK_{+})+\dot{h}e^{-g}(K_{-}-2fK_{3}+f^{2}K_{+})\right\}B(t) \\
&=&\left\{(\dot{f}-\dot{g}f+\dot{h}e^{-g}f^{2})K_{+}+(\dot{g}-2\dot{h}e^{-g}f)K_{3}+\dot{h}e^{-g}K_{-}\right\}B(t)
\end{eqnarray*}
where we have used relations
\begin{eqnarray*}
&&e^{fK_{+}}K_{3}e^{-fK_{+}}=K_{3}-fK_{+}, \\
&&e^{gK_{3}}K_{-}e^{-gK_{3}}=e^{-g}K_{-}\quad\mbox{and}\quad
e^{fK_{+}}K_{-}e^{-fK_{+}}=K_{-}-2fK_{3}+f^{2}K_{+}.
\end{eqnarray*}
The proof is easy. By comparing coefficients of $\dot{A}(t)$ and $\dot{B}(t)$ 
we have
\[
\left\{
\begin{array}{l}
\dot{h}e^{-g}=\mu, \\
\dot{g}-2\dot{h}e^{-g}f=-(\mu+\nu), \\
\dot{f}-\dot{g}f+\dot{h}e^{-g}f^{2}=\nu
\end{array}
\right.
\Longrightarrow 
\left\{
\begin{array}{l}
\dot{h}e^{-g}=\mu, \\
\dot{g}-2\mu f=-(\mu+\nu), \\
\dot{f}-\dot{g}f+\mu f^{2}=\nu
\end{array}
\right.
\Longrightarrow 
\left\{
\begin{array}{l}
\dot{h}e^{-g}=\mu, \\
\dot{g}-2\mu f=-(\mu+\nu), \\
\dot{f}+(\mu+\nu)f-\mu f^{2}=\nu.
\end{array}
\right.
\]
Note that the equation
\[
\dot{f}+(\mu+\nu)f-\mu f^{2}=\nu
\]
is a (famous) Riccati equation. If we can solve the equation then 
we obtain solutions like
\[
f\ \Longrightarrow\ g\ \Longrightarrow\ h.
\]
Unfortunately, it is not easy.  However there is an ansatz for 
the solution, $G$, $F$ and $E$. That is,
\[
f(t)=G(t),\quad g(t)=-2\log(F(t)),\quad h(t)=E(t)
\]
in (\ref{eq:a set of solutions}). To check these equations is left to 
readers (as a good exercise). From this 
\[
A(0)=B(0),\quad \dot{A}(0)=\dot{B}(0)\ \Longrightarrow\ 
A(t)=B(t)\quad \mbox{for all}\ \ t
\]
and we finally obtain the disentangling formula
\begin{equation}
\label{eq:}
e^{t\{\nu K_{+}+\mu K_{-}-(\mu+\nu)K_{3}\}}=
e^{G(t)K_{+}}e^{-2\log(F(t))K_{3}}e^{E(t)K_{-}}
\end{equation}
with (\ref{eq:a set of solutions}). 

Therefore (\ref{eq:formal-solution}) becomes
\[
\widehat{\rho}(t)=
e^{\frac{\mu-\nu}{2}t}
\exp\left(G(t)a^{\dagger}\otimes a^{\dagger}\right)
\exp\left(-\log(F(t))
    (N\otimes {\bf 1}+{\bf 1}\otimes N+{\bf 1}\otimes {\bf 1})\right)
\exp\left(E(t)a\otimes a\right)
\widehat{\rho}(0)
\]
with (\ref{eq:a set of solutions}). Some calculation by use of 
(\ref{eq:useful properties}) gives
\begin{eqnarray}
\label{eq:approximate-solution-4}
\widehat{\rho}(t)
=\hspace{-5mm}
&&\frac{e^{\frac{\mu-\nu}{2}t}}{F(t)}
\exp\left(G(t)a^{\dagger}\otimes a^{T}\right)
\left\{
\exp\left(-\log(F(t))N\right)\otimes 
\exp\left(-\log(F(t))N\right)^{T}
\right\}\times    \nonumber \\
&&\qquad \ \ \exp\left(E(t)a\otimes (a^{\dagger})^{T}\right)
\widehat{\rho}(0)
\end{eqnarray}
where we have used $N^{T}=N$ and $a^{\dagger}=a^{T}$.
By coming back to matrix form by use of (\ref{eq:well-known formulas}) 
like
\begin{eqnarray*}
\exp\left(E(t)a\otimes (a^{\dagger})^{T}\right)\widehat{\rho}(0)
&=&
\sum_{m=0}^{\infty}\frac{E(t)^{m}}{m!}\left(a\otimes(a^{\dagger})^{T}\right)^{m}\widehat{\rho}(0) \\
&=&
\sum_{m=0}^{\infty}\frac{E(t)^{m}}{m!}\left(a^{m}\otimes((a^{\dagger})^{m})^{T}\right)\widehat{\rho}(0)
\longrightarrow 
\sum_{m=0}^{\infty}
\frac{E(t)^{m}}{m!}a^{m}\rho(0)(a^{\dagger})^{m}
\end{eqnarray*}
we finally obtain
\begin{eqnarray}
\label{eq:matrix form}
&&\rho(t)=
\frac{e^{\frac{\mu-\nu}{2}t}}{F(t)}\times \nonumber \\
&&\sum_{n=0}^{\infty}
\frac{G(t)^{n}}{n!}(a^{\dagger})^{n}
\left[
\exp\left(-\log(F(t))N\right)
\left.
\left\{
\sum_{m=0}^{\infty}
\frac{E(t)^{m}}{m!}a^{m}\rho(0)(a^{\dagger})^{m}
\right\}
\right.
\exp\left(-\log(F(t))N\right)
\right]
a^{n}. \nonumber \\
&{}&
\end{eqnarray}
This form is very beautiful but complicated !

\subsection{General Solution}
Last, we treat the full equation (\ref{eq:quantum damped harmonic oscillator})
\[
\frac{\partial}{\partial t}\rho=-i\omega(a^{\dagger}a\rho-\rho a^{\dagger}a)
-
\frac{\mu}{2}
\left(a^{\dagger}a\rho+\rho a^{\dagger}a-2a\rho a^{\dagger}\right)
-
\frac{\nu}{2}
\left(aa^{\dagger}\rho+\rho aa^{\dagger}-2a^{\dagger}\rho{a}\right).
\]
From the lesson in the preceding subsection it is easy to rewrite 
this as
\begin{equation}
\label{eq:full-equation}
\frac{\partial}{\partial t}\widehat{\rho}
=
\left\{
-i\omega K_{0}+\nu K_{+}+\mu K_{-}-(\mu+\nu)K_{3}
+\frac{\mu-\nu}{2}{\bf 1}\otimes {\bf 1}
\right\}
\widehat{\rho}
\end{equation}
in terms of $K_{0}=N\otimes {\bf 1}-{\bf 1}\otimes N$ (note that 
$N^{T}=N$). Then it is easy to see
\begin{equation}
\label{eq:essential relations}
[K_{0},K_{+}]=[K_{0},K_{3}]=[K_{0},K_{-}]=0
\end{equation}
from (\ref{eq:K-generators}), which is left to readers. 
That is, $K_{0}$ commutes with all $\{K_{+}, K_{3}, K_{-}\}$. 
Therefore 
\begin{eqnarray*}
\widehat{\rho}(t)
&=&
e^{-i\omega t K_{0}}
e^{t\{\nu K_{+}+\mu K_{-}-(\mu+\nu) K_{3}
+\frac{\mu-\nu}{2}{\bf 1}\otimes {\bf 1}\}}
\widehat{\rho}(0) \\
&=&
e^{\frac{\mu-\nu}{2}t}
\exp\left(-i\omega t K_{0}\right)
\exp\left(G(t)K_{+}\right)
\exp\left(-2\log(F(t))K_{3}\right)
\exp\left(E(t)K_{-}\right)
\widehat{\rho}(0) \\
&=&
e^{\frac{\mu-\nu}{2}t}
\exp\left(G(t)K_{+}\right)
\exp\left(\{-i\omega t K_{0}-2\log(F(t))K_{3}\}\right)
\exp\left(E(t)K_{-}\right)
\widehat{\rho}(0),
\end{eqnarray*}
so that the general solution that we are looking for is 
just given by
\begin{eqnarray}
\label{eq:final form}
&&\rho(t)=
\frac{e^{\frac{\mu-\nu}{2}t}}{F(t)}
\sum_{n=0}^{\infty}
\frac{G(t)^{n}}{n!}(a^{\dagger})^{n}
[
\exp\left(\{-i\omega t-\log(F(t))\}N\right)\times 
\nonumber \\
&&\qquad\qquad\qquad\quad
\left\{
\sum_{m=0}^{\infty}
\frac{E(t)^{m}}{m!}a^{m}\rho(0)(a^{\dagger})^{m}
\right\}
\exp\left(\{i\omega t-\log(F(t))\}N\right)
]
a^{n}
\end{eqnarray}
by use of (\ref{eq:approximate-solution-4}) and 
(\ref{eq:matrix form}).

Particularly, if $\nu=0$ then
\begin{eqnarray*}
&&E(t)
=\frac{2\sinh\left(\frac{\mu}{2}t\right)}
{\cosh\left(\frac{\mu}{2}t\right)+\sinh\left(\frac{\mu}{2}t\right)}
=1-e^{-\mu t}, \\
&&F(t)
=\cosh\left(\frac{\mu}{2}t\right)+\sinh\left(\frac{\mu}{2}t\right)
=e^{\frac{\mu}{2}t}, \\
&&G(t)=0
\end{eqnarray*}
from (\ref{eq:a set of solutions}), so that we have
\begin{equation}
\label{eq:corollary}
\rho(t)=
e^{-\left(\frac{\mu}{2}+i\omega\right)tN}
\left\{
\sum_{m=0}^{\infty}
\frac{\left(1-{e}^{-\mu t}\right)^{m}}{m!}a^{m}\rho(0)(a^{\dagger})^{m}
\right\}
e^{-\left(\frac{\mu}{2}-i\omega\right)tN}
\end{equation}
from (\ref{eq:final form}).

\section{Quantum Counterpart}
In this section we explicitly calculate $\rho(t)$ for the initial 
value $\rho(0)$ given in the following.

\subsection{Case of $\rho(0)=|{0}\rangle\langle{0}|$}
Noting $a|{0}\rangle=0\ (\Leftrightarrow 0=\langle{0}|a^{\dagger})$,  
this case is very easy and we have
\begin{equation}
\label{eq:special case}
\rho(t)
=
\frac{e^{\frac{\mu-\nu}{2}t}}{F(t)}
\sum_{n=0}^{\infty}\frac{G(t)^{n}}{n!}(a^{\dagger})^{n}\ket{0}\bra{0}a^{n}
=
\frac{e^{\frac{\mu-\nu}{2}t}}{F(t)}
\sum_{n=0}^{\infty}G(t)^{n}\ket{n}\bra{n}
=
\frac{e^{\frac{\mu-\nu}{2}t}}{F(t)}e^{\log G(t) N}  
\end{equation}
because the number operator $N\ (=a^{\dagger}a)$ is written as
\[
N=\sum_{n=0}^{\infty}n\ket{n}\bra{n}
\ \Longrightarrow\ N\ket{n}=n\ket{n}
\]
, see for example (\ref{eq:creation-annihilation}). To check the 
last equality of (\ref{eq:special case}) is left to readers. 
Moreover, $\rho(t)$ can be written as
\begin{equation}
\label{eq:special case-modified}
\rho(t)
=(1-G(t))e^{\log G(t) N}=e^{\log(1-G(t))}e^{\log G(t) N},
\end{equation}
see the next subsection.

\subsection{Case of $\rho(0)=|{\alpha}\rangle\langle{\alpha}|\ (\alpha\in {\bf C})$}
Remind that $|{\alpha}\rangle$ is a coherent state given 
by (\ref{eq:three equations}) ($a|{\alpha}\rangle=\alpha|{\alpha}\rangle 
\Leftrightarrow \langle{\alpha}|a^{\dagger}=\langle{\alpha}|\bar{\alpha}$). 
First of all let us write down the result :
\begin{equation}
\label{eq:quantum counterpart}
\rho(t)
=
e^{|\alpha|^{2}e^{-(\mu-\nu)t}\log G(t)+\log(1-G(t))}
\exp\left\{-\log G(t)
\left(
\alpha e^{-\left(\frac{\mu-\nu}{2}+i\omega\right)t}a^{\dagger}+
\bar{\alpha}e^{-\left(\frac{\mu-\nu}{2}-i\omega\right)t}a-N
\right)
 \right\}
\end{equation}
with $G(t)$ in (\ref{eq:a set of solutions}). 
Here we again meet a term like (\ref{eq:approximate solution})
\[
\alpha e^{-\left(\frac{\mu-\nu}{2}+i\omega\right)t}a^{\dagger}+
\bar{\alpha}e^{-\left(\frac{\mu-\nu}{2}-i\omega\right)t}a
\]
with $\lambda=\frac{\mu-\nu}{2}$. 

Therefore, (\ref{eq:quantum counterpart}) is just our quantum 
counterpart of the classical damped harmonic oscillator.  

The proof is divided into four parts.

\vspace{5mm}\noindent
[First Step]\quad From (\ref{eq:final form}) it is easy to see
\[
\sum_{m=0}^{\infty}\frac{E(t)^{m}}{m!}a^{m}\ket{\alpha}\bra{\alpha}
(a^{\dagger})^{m}
=
\sum_{m=0}^{\infty}\frac{E(t)^{m}}{m!}{\alpha}^{m}\ket{\alpha}\bra{\alpha}
{\bar{\alpha}}^{m}
=
\sum_{m=0}^{\infty}\frac{(E(t)|\alpha|^{2})^{m}}{m!}\ket{\alpha}\bra{\alpha}
=
e^{E(t)|\alpha|^{2}}\ket{\alpha}\bra{\alpha}.
\]

\vspace{5mm}\noindent
[Second Step]\quad From (\ref{eq:final form}) we must calculate 
the term
\[
e^{\gamma N}\ket{\alpha}\bra{\alpha}e^{\bar{\gamma}N}
=
e^{\gamma N}e^{\alpha a^{\dagger}-\bar{\alpha}a}\ket{0}
\bra{0}e^{-(\alpha a^{\dagger}-\bar{\alpha}a)}e^{\bar{\gamma}N}
\]
where $\gamma=-i\omega t-\log(F(t))$ (note $\bar{\gamma}\ne -\gamma$). 
It is easy to see
\[
e^{\gamma N}e^{\alpha a^{\dagger}-\bar{\alpha}a}\ket{0}
=
e^{\gamma N}e^{\alpha a^{\dagger}-\bar{\alpha}a}e^{-\gamma N}
e^{\gamma N}\ket{0}
=
e^{e^{\gamma N}(\alpha a^{\dagger}-\bar{\alpha}a)e^{-\gamma N}}
\ket{0}
=
e^{\alpha e^{\gamma}a^{\dagger}-\bar{\alpha}e^{-\gamma}a}\ket{0}
\]
where we have used
\[
e^{\gamma N}a^{\dagger}e^{-\gamma N}=e^{\gamma}a^{\dagger}
\quad \mbox{and}\quad
e^{\gamma N}ae^{-\gamma N}=e^{-\gamma}a.
\]
The proof is easy and left to readers. 
Therefore, by use of Baker--Campbell--Hausdorff formula 
(\ref{eq:special B-C-H}) two times
\begin{eqnarray*}
e^{\alpha e^{\gamma}a^{\dagger}-\bar{\alpha}e^{-\gamma}a}\ket{0}
&=&
e^{-\frac{|\alpha|^{2}}{2}}
e^{\alpha e^{\gamma}a^{\dagger}}e^{-\bar{\alpha}e^{-\gamma}a}\ket{0}
=
e^{-\frac{|\alpha|^{2}}{2}}e^{\alpha e^{\gamma}a^{\dagger}}\ket{0} \\
&=&
e^{-\frac{|\alpha|^{2}}{2}}e^{\frac{|\alpha|^{2}}{2}e^{\gamma +\bar{\gamma}}}
e^{\alpha e^{\gamma}a^{\dagger}-\bar{\alpha}e^{\bar{\gamma}}a}\ket{0}
=
e^{-\frac{|\alpha|^{2}}{2}(1-e^{\gamma +\bar{\gamma}})}\ket{\alpha e^{\gamma}}
\end{eqnarray*}
and we obtain
\[
e^{\gamma N}\ket{\alpha}\bra{\alpha}e^{\bar{\gamma}N}
=
e^{-|\alpha|^{2}(1-e^{\gamma +\bar{\gamma}})}
\ket{\alpha e^{\gamma}}\bra{\alpha e^{\gamma}}
\]
with $\gamma=-i\omega t-\log(F(t))$.

\vspace{5mm}\noindent
[Third Step]\quad Under two steps above the equation 
(\ref{eq:final form}) becomes 
\[
\rho(t)=
\frac{
e^{\frac{\mu-\nu}{2}t}
e^{|\alpha|^{2}(E(t)-1+e^{\gamma +\bar{\gamma}})}}{F(t)}
\sum_{n=0}^{\infty}
\frac{G(t)^{n}}{n!}(a^{\dagger})^{n}
\ket{\alpha e^{\gamma}}\bra{\alpha e^{\gamma}}a^{n}.
\]
For simplicity we set $z=\alpha e^{\gamma}$ and calculate 
the term
\[
(\sharp)=\sum_{n=0}^{\infty}
\frac{G(t)^{n}}{n!}(a^{\dagger})^{n}\ket{z}\bra{z}a^{n}.
\]
Since $\ket{z}=e^{-|z|^{2}/2}e^{za^{\dagger}}\ket{0}$ we have
\begin{eqnarray*}
(\sharp)
&=&
e^{-|z|^{2}}
\sum_{n=0}^{\infty}
\frac{G(t)^{n}}{n!}(a^{\dagger})^{n}e^{za^{\dagger}}\ket{0}
\bra{0}e^{\bar{z}a}a^{n} \\
&=&
e^{-|z|^{2}}
e^{za^{\dagger}}
\left\{
\sum_{n=0}^{\infty}
\frac{G(t)^{n}}{n!}(a^{\dagger})^{n}\ket{0}\bra{0}a^{n}
\right\}
e^{\bar{z}a} \\
&=&
e^{-|z|^{2}}
e^{za^{\dagger}}
\left\{\sum_{n=0}^{\infty}G(t)^{n}\ket{n}\bra{n}\right\}
e^{\bar{z}a} \\
&=&
e^{-|z|^{2}}e^{za^{\dagger}}e^{\log G(t)N}e^{\bar{z}a}
\end{eqnarray*}
by (\ref{eq:special case}). 
Namely, this form is a kind of disentangling formula, so 
we want to restore an entangling formula.

For the purpose we use the {\bf disentangling formula}
\begin{equation}
\label{eq:disentangling formula-1}
e^{\alpha a^{\dagger}+\beta a+\gamma N}
=
e^{\alpha\beta\frac{e^{\gamma}-(1+\gamma)}{\gamma^{2}}}
e^{\alpha\frac{e^{\gamma}-1}{\gamma}a^{\dagger}}
e^{\gamma N}
e^{\beta\frac{e^{\gamma}-1}{\gamma}a}
\end{equation}
where $\alpha$, $\beta$, $\gamma$ are usual numbers. 
The proof is given in the fourth step. From this it is easy to see
\begin{equation}
\label{eq:disentangling formula-2}
e^{ua^{\dagger}}e^{vN}e^{wa}
=
e^{-\frac{uw(e^{v}-(1+v))}{(e^{v}-1)^{2}}}
e^{\frac{uv}{e^{v}-1}a^{\dagger}+\frac{vw}{e^{v}-1}a+vN}.
\end{equation}
Therefore ($u\rightarrow z,\ v\rightarrow \log G(t),\ w\rightarrow \bar{z}$)
\[
(\sharp)=e^{-|z|^{2}}
e^{\frac{|z|^{2}(1+\log G(t)-G(t))}{(1-G(t))^{2}}}
e^{
\frac{\log G(t)}{G(t)-1}za^{\dagger}+
\frac{\log G(t)}{G(t)-1}\bar{z}a+
\log G(t)N
},
\]
so by noting
\[
z=\alpha e^{\gamma}=\alpha \frac{e^{-i\omega t}}{F(t)}
\quad \mbox{and}\quad
|z|^{2}
=|\alpha|^{2}e^{\gamma + \bar{\gamma}}
=|\alpha|^{2}\frac{1}{F(t)^{2}}
\]
we have
\begin{eqnarray*}
\rho(t)
&=&
\frac{e^{\frac{\mu-\nu}{2}t}}{F(t)}
e^{|\alpha|^{2}(E(t)-1)}
e^{|\alpha|^{2}\frac{1+\log G(t)-G(t)}{F(t)^{2}(1-G(t))^{2}}}
e^{
\frac{\log G(t)}{F(t)(G(t)-1)}\alpha e^{-i\omega t}a^{\dagger}+
\frac{\log G(t)}{F(t)(G(t)-1)}\bar{\alpha} e^{i\omega t}a+
\log G(t)N
}  \\
&=&
\frac{e^{\frac{\mu-\nu}{2}t}}{F(t)}
e^{|\alpha|^{2}\left\{E(t)-1+
    \frac{1+\log G(t)-G(t)}{F(t)^{2}(1-G(t))^{2}}\right\}}
e^{
\frac{\log G(t)}{F(t)(G(t)-1)}\alpha e^{-i\omega t}a^{\dagger}+
\frac{\log G(t)}{F(t)(G(t)-1)}\bar{\alpha} e^{i\omega t}a+
\log G(t)N 
}.
\end{eqnarray*}

By the way, from (\ref{eq:a set of solutions}) 
\[
G(t)-1=-\frac{e^{\frac{\mu-\nu}{2}t}}{F(t)},\quad
\frac{1}{F(t)(G(t)-1)}=-e^{-\frac{\mu-\nu}{2}t},\quad
E(t)-1=-\frac{e^{-\frac{\mu-\nu}{2}t}}{F(t)}
\]
and
\begin{eqnarray*}
\frac{1-G(t)+\log G(t)}{F(t)^{2}(G(t)-1)^{2}}
&=&
e^{-(\mu-\nu)t}
\left\{
\frac{e^{\frac{\mu-\nu}{2}t}}{F(t)}+\log G(t)
\right\} \\
&=&
\frac{e^{-\frac{\mu-\nu}{2}t}}{F(t)}+e^{-(\mu-\nu)t}\log G(t) \\
&=&
-(E(t)-1)+e^{-(\mu-\nu)t}\log G(t)
\end{eqnarray*}
we finally obtain
\begin{eqnarray*}
\rho(t)
&=&
(1-G(t))e^{|\alpha|^{2}e^{-(\mu-\nu)t}\log G(t)}
e^{-\log G(t)
\left\{
\alpha e^{-i\omega t}e^{-\frac{\mu-\nu}{2}t}a^{\dagger}+
\bar{\alpha} e^{i\omega t}e^{-\frac{\mu-\nu}{2}t}a-
N
\right\}
} \\
&=&
e^{|\alpha|^{2}e^{-(\mu-\nu)t}\log G(t)+\log (1-G(t))}
e^{-\log G(t)
\left\{
\alpha e^{-\left(\frac{\mu-\nu}{2}+i\omega\right)t}a^{\dagger}+
\bar{\alpha} e^{-\left(\frac{\mu-\nu}{2}-i\omega\right)t}a-
N
\right\}
}.
\end{eqnarray*}

\vspace{5mm}\noindent
[Fourth Step]\quad In last, let us give the proof to the disentangling 
formula (\ref{eq:disentangling formula-1}) because it is not so popular 
as far as we know. 
From (\ref{eq:disentangling formula-1}) 
\begin{eqnarray*}
\alpha a^{\dagger}+\beta a+\gamma N
&=&
\gamma a^{\dagger}a+\alpha a^{\dagger}+\beta a \\
&=&
\gamma \left\{
\left(a^{\dagger}+\frac{\beta}{\gamma}\right)
\left(a+\frac{\alpha}{\gamma}\right)
-\frac{\alpha\beta}{\gamma^{2}}\right\} \\
&=&
\gamma 
\left(a^{\dagger}+\frac{\beta}{\gamma}\right)
\left(a+\frac{\alpha}{\gamma}\right)
-\frac{\alpha\beta}{\gamma}
\end{eqnarray*}
we have
\begin{eqnarray*}
e^{\alpha a^{\dagger}+\beta a+\gamma N}
&=&
e^{-\frac{\alpha\beta}{\gamma}}
e^{\gamma 
\left(a^{\dagger}+\frac{\beta}{\gamma}\right)
\left(a+\frac{\alpha}{\gamma}\right)} \\
&=&
e^{-\frac{\alpha\beta}{\gamma}}
e^{\frac{\beta}{\gamma}a}
e^{\gamma a^{\dagger}
\left(a+\frac{\alpha}{\gamma}\right)} 
e^{-\frac{\beta}{\gamma}a}  \\
&=&
e^{-\frac{\alpha\beta}{\gamma}}
e^{\frac{\beta}{\gamma}a}
e^{-\frac{\alpha}{\gamma}a^{\dagger}}
e^{\gamma a^{\dagger}a}
e^{\frac{\alpha}{\gamma}a^{\dagger}}
e^{-\frac{\beta}{\gamma}a}.
\end{eqnarray*}
Then, careful calculation gives 
the disentangling formula (\ref{eq:disentangling formula-1}) 
($N=a^{\dagger}a$)
\begin{eqnarray*}
e^{-\frac{\alpha\beta}{\gamma}}
e^{\frac{\beta}{\gamma}a}
e^{-\frac{\alpha}{\gamma}a^{\dagger}}
e^{\gamma N}
e^{\frac{\alpha}{\gamma}a^{\dagger}}
e^{-\frac{\beta}{\gamma}a}
&=&
e^{-\frac{\alpha\beta}{\gamma}}
e^{-\frac{\alpha\beta}{\gamma^{2}}}
e^{-\frac{\alpha}{\gamma}a^{\dagger}}
e^{\frac{\beta}{\gamma}a}
e^{\gamma N}
e^{\frac{\alpha}{\gamma}a^{\dagger}}
e^{-\frac{\beta}{\gamma}a}  \\
&=&
e^{-(\frac{\alpha\beta}{\gamma}+
    \frac{\alpha\beta}{\gamma^{2}})}
e^{-\frac{\alpha}{\gamma}a^{\dagger}}
e^{\frac{\beta}{\gamma}a}
e^{\gamma N}
e^{\frac{\alpha}{\gamma}a^{\dagger}}
e^{-\frac{\beta}{\gamma}a}  \\
&=&
e^{-(\frac{\alpha\beta}{\gamma}+
    \frac{\alpha\beta}{\gamma^{2}})}
e^{-\frac{\alpha}{\gamma}a^{\dagger}}
e^{\gamma N}
e^{\frac{\beta}{\gamma}e^{\gamma}a}
e^{\frac{\alpha}{\gamma}a^{\dagger}}
e^{-\frac{\beta}{\gamma}a} \\
&=&
e^{-(\frac{\alpha\beta}{\gamma}+
    \frac{\alpha\beta}{\gamma^{2}})
   +\frac{\alpha\beta}{\gamma^{2}}e^{\gamma}}
e^{-\frac{\alpha}{\gamma}a^{\dagger}}
e^{\gamma N}
e^{\frac{\alpha}{\gamma}a^{\dagger}}
e^{\frac{\beta}{\gamma}e^{\gamma}a}
e^{-\frac{\beta}{\gamma}a}  \\
&=&
e^{\alpha\beta\frac{e^{\gamma}-1-\gamma}{\gamma^{2}}}
e^{-\frac{\alpha}{\gamma}a^{\dagger}}
e^{\frac{\alpha}{\gamma}e^{\gamma}a^{\dagger}}
e^{\gamma N}
e^{\beta\frac{e^{\gamma}-1}{\gamma}a} \\
&=&
e^{\alpha\beta\frac{e^{\gamma}-1-\gamma}{\gamma^{2}}}
e^{\alpha\frac{e^{\gamma}-1}{\gamma}a^{\dagger}}
e^{\gamma N}
e^{\beta\frac{e^{\gamma}-1}{\gamma}a}
\end{eqnarray*}
by use of some commutation relations
\[
e^{sa}e^{ta^{\dagger}}=e^{st}e^{ta^{\dagger}}e^{sa},
\quad
e^{sa}e^{tN}=e^{tN}e^{se^{t}a},
\quad
e^{tN}e^{sa^{\dagger}}=e^{se^{t}a^{\dagger}}e^{tN}.
\]
The proof is simple. For example, 
\[
e^{sa}e^{ta^{\dagger}}=e^{sa}e^{ta^{\dagger}}e^{-sa}e^{sa}=
e^{te^{sa}a^{\dagger}e^{-sa}}e^{sa}=
e^{t(a^{\dagger}+s)}e^{sa}=
e^{st}e^{ta^{\dagger}}e^{sa}.
\]
The remainder is left to readers.

\vspace{3mm}
We finished the proof. 
The formula (\ref{eq:quantum counterpart}) is both compact and 
clear-cut and has not been known as far as we know. 
See \cite{BP} and \cite{WS} for some applications.

\vspace{5mm}
In last, let us present a challenging problem. A squeezed state 
$\ket{\beta}\ (\beta\in \fukuso)$ is defined as
\begin{equation}
\label{eq:squeezed state}
\ket{\beta}=
e^{\frac{1}{2}\left(\beta (a^{\dagger})^{2}-\bar{\beta}a^{2}\right)}\ket{0}.
\end{equation}
See for example \cite{KF-1}. For the initial value 
\
$
\rho(0)=\ket{\beta}\bra{\beta}
$ 
\
we want to calculate $\rho(t)$ in (\ref{eq:final form}) like in the text. 
However, we cannot sum up it in a compact form like 
(\ref{eq:quantum counterpart}) at the present time, so we propose 
the problem,

\noindent
{\bf Problem}\ \ sum up $\rho(t)$ in a compact form.

\section{Concluding Remarks}
In this chapter we treated the quantum damped harmonic oscillator, 
and studied mathematical structure of the model, and constructed 
general solution with any initial condition, and gave a quantum 
counterpart in the case of taking coherent state as an initial 
condition. It is in my opinion perfect. 

However, readers should pay attention to the fact that this is 
not a goal but a starting point. Our real target is to construct 
general theory of {\bf Quantum Mechanics with Dissipation}.

In the papers \cite{FS-1} and \cite{FS-2} (see also \cite{JC}) 
we studied a more realistic model entitled ``Jaynes--Cummings 
Model with Dissipation" and constructed some approximate solutions 
under any initial condition.  In the paper \cite{KF-2} we studied 
``Superluminal Group Velocity of Neutrinos" from the point of view  
of Quantum Mechanics with Dissipation.

Unfortunately, there is no space to introduce them. It is a good 
challenge for readers to read them carefully and attack 
the problems.

\vspace{10mm}\noindent
{\em Acknowledgment.}\\
I would like to thank Ryu Sasaki and Tatsuo Suzuki for 
their helpful comments and suggestions.


\end{document}